\documentstyle[epsfig]{aipproc}
\begin{document}
\begin{flushright}
UB-HET-97-04\\
December 1997\\[0.5cm]
\end{flushright}
\title{Measuring Trilinear Gauge Boson Couplings at Hadron and Lepton  
Colliders
\thanks{Talk given at the {\sl Workshop on Physics at the First 
Muon Collider and the Front End of a Muon Collider}, Fermilab, November~6 
--~9, 1997, to appear in the Proceedings.}}

\author{U. Baur\thanks{Research supported by National Science Foundation
Grant No.~PHY9600770.}}
\address{Department of Physics\\
State University of New York at Buffalo\\
Buffalo, NY 14260}

\maketitle

\begin{abstract}
We discuss the measurement of the $WWV$ ($V=\gamma,\, Z$) gauge boson
couplings in present and future collider experiments. The major goals of
such experiments will be the confirmation of the Standard Model (SM)
predictions and the search for signals of new physics. 
The present limits on these couplings from Tevatron and LEP2 experiments
as well as the expectations from future hadron and lepton collider
experiments are summarized. We also study  the impact of initial state 
radiation on the sensitivity limits which can be achieved at the NLC and a
$\mu^+\mu^-$ collider. 
\end{abstract}

\section*{Introduction}
Over the last seven years $e^+e^-$ collision experiments at LEP and at
the SLAC
linear collider have beautifully confirmed the predictions of the Standard 
Model (SM). At present experiment and theory agree at the 0.1 --~1\% level 
in the determination of the vector boson couplings to the various
fermions~\cite{ward}, which may rightly be considered a confirmation 
of the 
gauge boson nature of the $W$ and the~$Z$. On the other hand, the most direct 
consequences of the $SU(2)_L\times U(1)_Y$ gauge symmetry, the non-abelian 
self-couplings of the $W$, $Z$, and photon, are known with much less
experimental precision.

A direct measurement of these vector boson couplings is 
possible in hadron and lepton collider experiments, in particular via pair 
production processes like $e^+e^- \to W^+W^-$ and  $q \bar q \to
W^+W^-,\; W\gamma,\; WZ$. The first and 
major goal of such experiments will be a confirmation of the SM
predictions. A precise and direct measurement of the
trilinear couplings of the electroweak vector bosons
and the demonstration that they agree with the SM would 
beautifully corroborate spontaneously broken, non-abelian gauge
theories as the basic theoretical structure describing the fundamental 
interactions of nature. At the same time, such measurements may be used
to probe for new physics. However, if the energy scale of the new
physics responsible for the non-standard gauge boson couplings is $\sim
1$~TeV, these anomalous couplings are expected to be no larger than
${\cal O}(10^{-2})$~\cite{aihara}. 

In the following, we present an overview of how trilinear gauge boson
couplings are measured in collider experiments.
For simplicity, we shall restrict our discussion to the $WWV$
($V=\gamma,\,Z$) couplings; $Z\gamma V$ couplings are not discussed. 
Analogous to the introduction of arbitrary vector and axial vector 
couplings $g_V$ and $g_A$ for the coupling of gauge bosons to 
fermions, the measurement of the $WWV$ couplings can be made quantitative by 
introducing a more general $WWV$ vertex. For our discussion of experimental 
sensitivities we shall use a parameterization in terms of the 
phenomenological effective Lagrangian~\cite{HPZH}
\begin{eqnarray} \label{LeffWWV}
i{\cal L}_{eff}^{WWV} & = & g_{WWV}\, \Bigl[ g_1^V\left(
W^{\dagger}_{\mu\nu}W^{\mu}-W^{\dagger\, \mu}W_{\mu\nu}\right)V^{\nu} +
\kappa_V\,  W^{\dagger}_{\mu}W_{\nu}V^{\mu\nu} + \\ \nonumber
& & {\lambda_V\over m_W^2}\,
W^{\dagger}_{\rho\mu}{W^{\mu}}_{\nu}V^{\nu\rho} \Bigr]\; . 
\end{eqnarray}
Here the overall couplings are defined as $g_{WW\gamma}=e$ and
$g_{WWZ}= e \cot\theta_W$, $W_{\mu\nu}=\partial_\mu W_\nu - \partial_\nu
W_\mu$, and $V_{\mu\nu}=\partial_\mu V_\nu - \partial_\nu V_\mu$. Within 
the SM, at tree level, the couplings are given by $g_1^Z = g_1^\gamma = 
\kappa_Z = \kappa_\gamma = 1,\; \lambda_Z = \lambda_\gamma = 0$. For 
on-shell photons, $g_1^\gamma=1$ is
fixed by electromagnetic gauge invariance; $g_1^Z$ may, 
however, differ from its SM value. Deviations are given by the
anomalous couplings $\Delta g_1^Z \equiv (g_1^Z - 1)$, $\Delta 
\kappa_\gamma \equiv (\kappa_\gamma-1)$, $\Delta\kappa_Z \equiv (\kappa_Z-1)$, 
$\lambda_\gamma$, and $\lambda_Z$. 

The effective Lagrangian of Eq.~(\ref{LeffWWV}) parameterizes the
most general 
Lorentz invariant and $C$ and $P$ conserving $WWV$ vertex which can be 
observed in processes where the vector bosons couple to effectively massless
fermions. If $C$ or $P$ violating couplings are allowed, four additional
couplings, $g_4^V$, $g_5^V$, $\tilde\kappa_V$ and $\tilde\lambda_V$,
appear in the
effective Lagrangian~\cite{HPZH} and they all vanish in the SM, at tree
level. For simplicity, these couplings are not considered in this
report. 

The terms in ${\cal L}_{eff}^{WW\gamma}$ 
correspond to the lowest order terms in a multipole expansion of the 
$W-$photon interactions, the charge $Q_W$, the magnetic dipole 
moment $\mu_W$, and the electric quadrupole moment $q_W$  of the
$W^+$~\cite{AR}:
\begin{eqnarray}\label{EQ:multipole}
Q_W & = & e g_1^\gamma\; , \\
\mu_W & = & {e \over 2m_W} 
\left(g_1^\gamma + \kappa_\gamma + \lambda_\gamma\right) \; , \\
q_W & = & -{e \over m_W^2} \left(\kappa_\gamma-\lambda_\gamma\right) \; . 
\end{eqnarray}

Terms with higher derivatives are equivalent to a dependence of the couplings 
on the vector boson momenta and thus merely lead to a form-factor 
behaviour of these couplings. Since a constant 
anomalous coupling would lead 
to unitarity violation at high energies~\cite{joglekar} such a form factor 
behaviour is a feature of any model of anomalous couplings. When studying 
$W^+W^-$ production at a lepton collider at fixed $q^2=s$ this form 
factor behaviour is of no consequence. 
Weak boson pair production at hadron colliders, however, probes the
gauge boson couplings  
over a large $q^2$ range and is very sensitive to the fall-off of 
anomalous couplings which necessarily happens once the threshold of new 
physics is crossed.
Not taking this cutoff into account results in unphysically large cross 
sections at high energy (which violate unitarity) and thus leads to a 
substantial overestimate of experimental sensitivities. In the following
we will assume a simple dipole behaviour, {\it e.g.}
\begin{equation}\label{FF}
\Delta\kappa_V(q^2) = { \Delta\kappa_V^0 \over (1+q^2/\Lambda_{FF}^2)^2 }\; ,
\end{equation}
and similarly for the other couplings. Here, $\Lambda_{FF}$ is the form
factor scale which is a function of the scale of new physics,
$\Lambda$. Due to the form factor behaviour of the anomalous couplings, the
experimental limits extracted from hadron collider experiments
explicitly depend on $\Lambda_{FF}$. 

From the phenomenological effective Lagrangian [see Eq.~(\ref{LeffWWV})]
it is straightforward
to derive cross section formulas for the di-boson production processes, 
$q\bar q'\to W^\pm\gamma$, $W^\pm Z$, and $W^+W^-$ production in $q\bar
q$, $e^+e^-$ and $\mu^+\mu^-$ annihilation. While the SM contribution to
the di-boson 
amplitudes is bounded from above for fixed scattering angle $\Theta$,
the anomalous contributions rise without
limit as $\hat s$ increases, eventually violating unitarity. This is
the reason the anomalous couplings must show a form
factor behavior at very high energies. Anomalous gauge boson couplings 
also affect the angular
distributions of the produced vector bosons in a characteristic way (see
Ref.~\cite{HPZH}). In hadronic collisions, the 
transverse momentum distribution of the
vector boson should be particularly sensitive to non-standard
$WWV$ couplings. At lepton colliders, on the other hand, where the
center of mass energy
is fixed, angular distributions are more useful. Hadron and lepton
collider data thus thus yield complementary information on the nature of
the $WWV$ couplings~\cite{aihara}.

\section*{Present Limits on $WWV$ Couplings}
Presently, the most stringent limits on anomalous $WWV$ couplings come
from the D\O\ and CDF experiments at the Tevatron, and from the four LEP
experiments. While the di-boson analysis of D\O\ is fairly complete, the 
CDF collaboration has not presented final results from their analysis of
run~1b data yet.

\subsection*{Tevatron Results}
The Tevatron experiments obtained information on the structure of the
$WWV$ vertices from $W\gamma$ production with subsequent $W\to\ell\nu$
($\ell=e,\,\mu$) decay, $p\bar p\to W^+W^-\to\ell\bar\nu\bar\ell'\nu$
($\ell,\,\ell' = e,\,\mu$), and $WW/WZ$ production with one of the
vector bosons decaying leptonically and the other gauge boson decaying
into two jets. Limits were also obtained from a combined fit to the
three processes. 

$W(\to\ell\nu)\gamma$ candidates were selected from the inclusive
$e/\mu$ channel $W$ samples by requiring an isolated photon with high
transverse energy ($E_T$). The main background sources for $W\gamma$ production
are $W+$~jets production where one of the jets ``fakes'' a photon, and
$Z\gamma$ production with one of the leptons from the $Z$ decay being
undetected. The signal to background ratio is about 1 to 0.2 --~0.3. 

\begin{figure}[t] % fig 1
\centerline{\epsfig{file=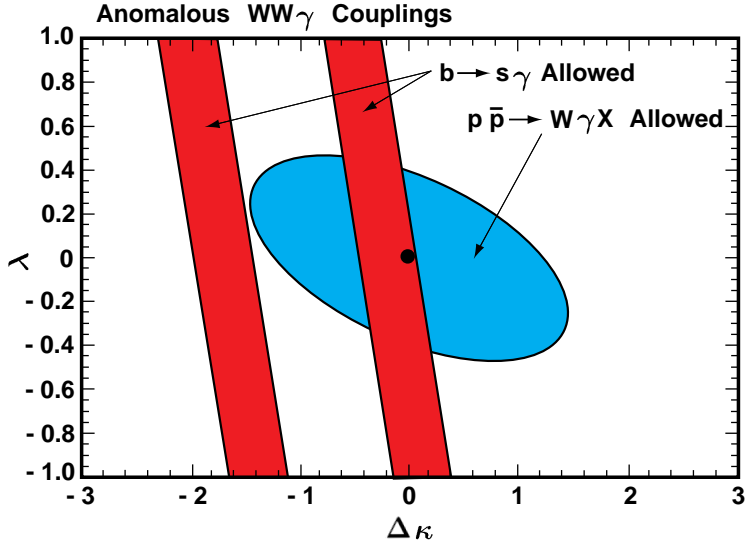,width=4.in}}
\vspace{10pt}
\caption{Present 95\% CL limits on anomalous $WW\gamma$ couplings from
$W\gamma$ and $b\to s\gamma$ data.}
\label{FIG:ONE}
\end{figure}

A detailed discussion of the CDF and D\O\ $W\gamma$ event selection can 
be found in Refs.~\cite{doug} and~\cite{dzerowg}. To set limits on the
$WW\gamma$ couplings $\Delta\kappa_\gamma$ and $\lambda_\gamma$, a
binned maximum likelihood fit to the photon $E_T$
spectrum was performed, using a Monte Carlo program based on the
calculation of Ref.~\cite{babe}, and a form factor scale of
$\Lambda_{FF}=1.5$~TeV. The D\O\ 95\% CL limit contour is shown
in Fig.~\ref{FIG:ONE}, together with the bands allowed by the 
CLEO~\cite{cleo} and ALEPH~\cite{aleph} $b\to s\gamma$ data.
The CDF and D\O\ 95\% CL limits on for $\Delta\kappa_\gamma$ and
$\lambda_\gamma$ are listed in Table~\ref{TAB:ONE}.
\begin{table}
\caption{95\% CL limits on $WW\gamma$ anomalous couplings from CDF and
D\O.}
\label{TAB:ONE}
\begin{tabular}{cccc}
 & $\lambda_\gamma^0=0$ & $\quad$ & $\Delta\kappa_\gamma^0=0$ \\[1.mm]
\tableline
CDF (67~pb$^{-1}$, preliminary)  & $-1.8<\Delta\kappa_\gamma^0<2.0$ & 
 $\quad$  & $-0.7<\lambda_\gamma^0<0.6$ \\[1.mm]
D\O\ (93~pb$^{-1}$)  & $-0.93<\Delta\kappa_\gamma^0<0.94$ & $\quad$ &
$-0.31<\lambda_\gamma^0<0.29$
\end{tabular}
\end{table}

Candidates for $WW\to$~dilepton production were selected by searching
for events with two isolated, high $E_T$ charged leptons ($e$ or $\mu$)
and large missing transverse energy, $E\llap/_T$. The main 
non-instrumental background in this
process is $\bar tt$ production which can be suppressed by applying a
cut on the hadronic energy in the event. The analysis is described in
detail in Ref.~\cite{wwdil}. From a fit to the electron $E_T$ and muon
$p_T$ distribution, the D\O\ Collaboration obtains the following
preliminary 95\% CL limits ($\int\! {\cal L}dt=97$~pb$^{-1}$)~\cite{taka}:
\begin{equation}
-0.62<\Delta\kappa^0<0.75\quad (\lambda^0=0), \quad\qquad -0.50
<\lambda^0 <0.56 \quad (\Delta\kappa^0=0), 
\end{equation}
where we have assumed $\Lambda_{FF}=1.5$~TeV, 
$\Delta\kappa^0=\Delta\kappa_\gamma^0=\Delta\kappa_Z^0$ and
$\lambda^0=\lambda_\gamma^0=\lambda_Z^0$. 

The $WW/WZ\to\ell\nu jj,\,\ell\ell jj$ data samples are extracted from 
inclusive $e/\mu$ $W/Z$ data, requiring two high $E_T$ jets in addition to
the $\ell\nu$ and $\ell^+\ell^-$ system. To reduce the enormous
background of QCD multijet and $W/Z+2$~jet production, a cut on the
di-jet invariant mass of $60~{\rm GeV}<m(jj)<110$~GeV ($50~{\rm
GeV}<m(jj)<110$~GeV) is imposed by CDF (D\O). More details of the
experimental analysis can be found in Refs.~\cite{wwsemi1} 
and~\cite{wwsemi2}. The 95\% CL
limits for $\Lambda_{FF}=2$~TeV obtained from the $W$ transverse
momentum distribution are listed in Table~\ref{TAB:TWO}.
\begin{table}
\caption{95\% CL limits on $WWV$ anomalous couplings from
$WW/WZ\to\ell\nu jj,\,\ell\ell jj$ production at the Tevatron, assuming 
$\Delta g_1^Z=0$,
$\Delta\kappa^0=\Delta\kappa_\gamma^0=\Delta\kappa_Z^0$, and
$\lambda^0=\lambda_\gamma^0=\lambda_Z^0$.}
\label{TAB:TWO}
\begin{tabular}{ccc}
 & $\lambda^0=0$ & $\Delta\kappa^0=0$ \\[1.mm]
\tableline
CDF (preliminary) & $-0.49<\Delta\kappa^0<0.54$ & $-0.35<\lambda^0<0.32$
\\[1.mm]
D\O\ & $-0.43<\Delta\kappa^0<0.59$ & $-0.33<\lambda^0<0.36$ 
\end{tabular}
\end{table}
A vanishing $WWZ$ vertex ($\Delta g_1^{Z0}=\kappa_Z^0=\lambda_Z^0=0$) is
excluded at the 99\% CL level by the experimental data.

The D\O\ Collaboration has also obtained preliminary limits on 
$\Delta\kappa^0=\Delta\kappa_\gamma^0=\Delta\kappa_Z^0$ and
$\lambda^0=\lambda_\gamma^0=\lambda_Z^0$ from a combined fit to the $W\gamma$,
$WW\to$~dilepton and $WW/WZ\to\ell\nu jj,\,\ell\ell jj$ data collected
in run~1~\cite{taka} ($\Lambda_{FF}=1.5$~TeV):
\begin{equation}
-0.33<\Delta\kappa^0<0.45\quad (\lambda^0=0), \quad\qquad -0.20
<\lambda^0 <0.20 \quad (\Delta\kappa^0=0). 
\end{equation}

\subsection*{LEP2 Results}
The four LEP experiments have recently presented~\cite{ward} measurements of
anomalous $WWV$ couplings parameters in $e^+e^-\to W^+W^-$ using the
1996 data set. A total integrated luminosity of approximately
10~pb$^{-1}$ was recorded at each, $\sqrt{s}=161$~GeV and 
$\sqrt{s}=172$~GeV, per experiment. The LEP experiments extract
limits on the following combination of anomalous $WWV$ couplings:
\begin{eqnarray}
\alpha_{W\phi} &=& \Delta g_1^Z\cos^2\theta_W, \\
\alpha_{B\phi} &=& \Delta\kappa_\gamma-\Delta g_1^Z\cos^2\theta_W, \\
\alpha_W       &=& \lambda_\gamma=\lambda_Z,
\label{EQ:ALPHA}
\end{eqnarray}
with the constraint $\Delta\kappa_Z=\Delta
g_1^Z-\Delta\kappa_\gamma\tan^2\theta_W$. Combined results from the four
experiments have been obtained by adding likelihood curves from each
experiment, taking both cross section and information on the angular
distributions of the final state fermions into account. The combined
95\% CL limits are~\cite{ward}
\begin{eqnarray}
-0.28 < & \alpha_{W\phi} & < 0.33 \qquad\qquad
(\alpha_{B\phi}=\alpha_W=0), \nonumber \\
-0.81 < & \alpha_{B\phi} & < 1.50 \qquad\qquad
(\alpha_{W\phi}=\alpha_W=0), \nonumber \\
-0.37 < & \alpha_W & < 0.68 \qquad\qquad (\alpha_{W\phi}=\alpha_{B\phi}=0).
\end{eqnarray}
Because of the different parameterization used by the LEP experiments,
it is difficult to compare these limits with the bounds extracted at the
Tevatron, except for $\alpha_W=\lambda_\gamma=\lambda_Z$. 

The OPAL Collaboration recently has published~\cite{opal} a measurement
of the $WWV$ couplings from the 1996 data which employs the
parameterization of Eq.~(\ref{LeffWWV}). In Table~\ref{TAB:THREE}, the
OPAL results are compared with the limits of the D\O\ $WW/WZ\to\ell\nu
jj$, $\ell\ell jj$ analysis ($\Lambda_{FF}=1.5$~TeV)~\cite{wwsemi2}.
\begin{table}
\caption{Comparison of the OPAL ($e^+e^-\to W^+W^-$, 1996 data) and D\O\
($WW/WZ\to\ell\nu jj$, $\ell\ell jj$) 95\% CL limits on $WWV$ anomalous 
couplings. Only one of the independent couplings is assumed to deviate
from the SM at a time. The OPAL limits have been corrected for form
factor effects ($\Lambda_{FF}=1.5$~TeV).}
\label{TAB:THREE}
\begin{tabular}{ccc}
 OPAL & $\qquad$ $\qquad$ & D\O\ \\[1.mm]
\tableline
$-0.77 < \Delta g_1^{Z0} < 0.79$ & $\qquad$ & $-0.64 < \Delta g_1^{Z0} < 0.89$
\\[1.mm]
$-0.92 < \Delta\kappa_\gamma^0=\Delta\kappa_Z^0 < 1.15$ & $\qquad$ $\qquad$ & 
$-0.48 < \Delta\kappa_\gamma^0=\Delta\kappa_Z^0 < 0.65$ \\[1.mm]
$-0.80 < \lambda_\gamma^0=\lambda_Z^0< 1.22$ & $\qquad$ $\qquad$ & 
$-0.36 < \lambda_\gamma^0=\lambda_Z^0< 0.39$
\end{tabular}
\end{table}
The limits obtained by the two experiments for $\Delta g_1^{Z0}$ are
similar. The D\O\ bounds for $\Delta\kappa_\gamma^0$ and $\lambda_\gamma^0$
are about a factor~2 to~3 better than the current OPAL limits. 

\section*{Future Limits: LEP2, Tevatron and LHC}
The bounds on anomalous gauge boson couplings from LEP2 are expected to
improve rapidly. In 1997, about 55~pb$^{-1}$ were recorded by each of 
the four experiments at a center of mass energy of $\sqrt{s}=183$~GeV. 
This should result in bounds
on non-standard $WWV$ couplings which are at least a factor~2 to~3 better than
the present LEP2 limits. Ultimately one hopes to collect 500~pb$^{-1}$ per
experiment and achieve a precision of 0.02~-- 0.1~\cite{lep2rep}. 

\begin{figure}[t] % fig 2
\begin{tabular}{ll}
\epsfig{file=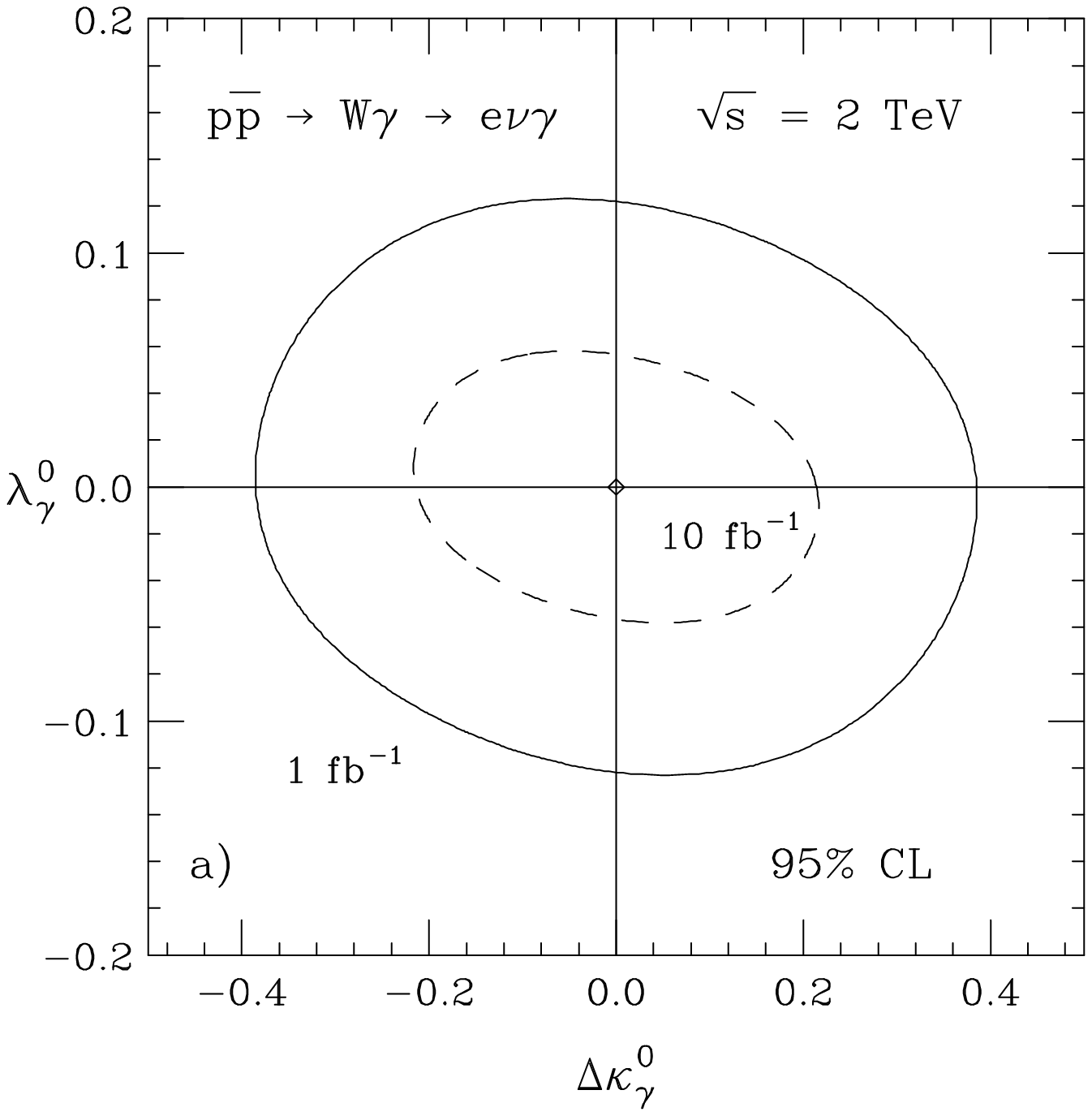,width=2.5in} & \hspace{-6pt}
%\vspace{10pt} 
\epsfig{file=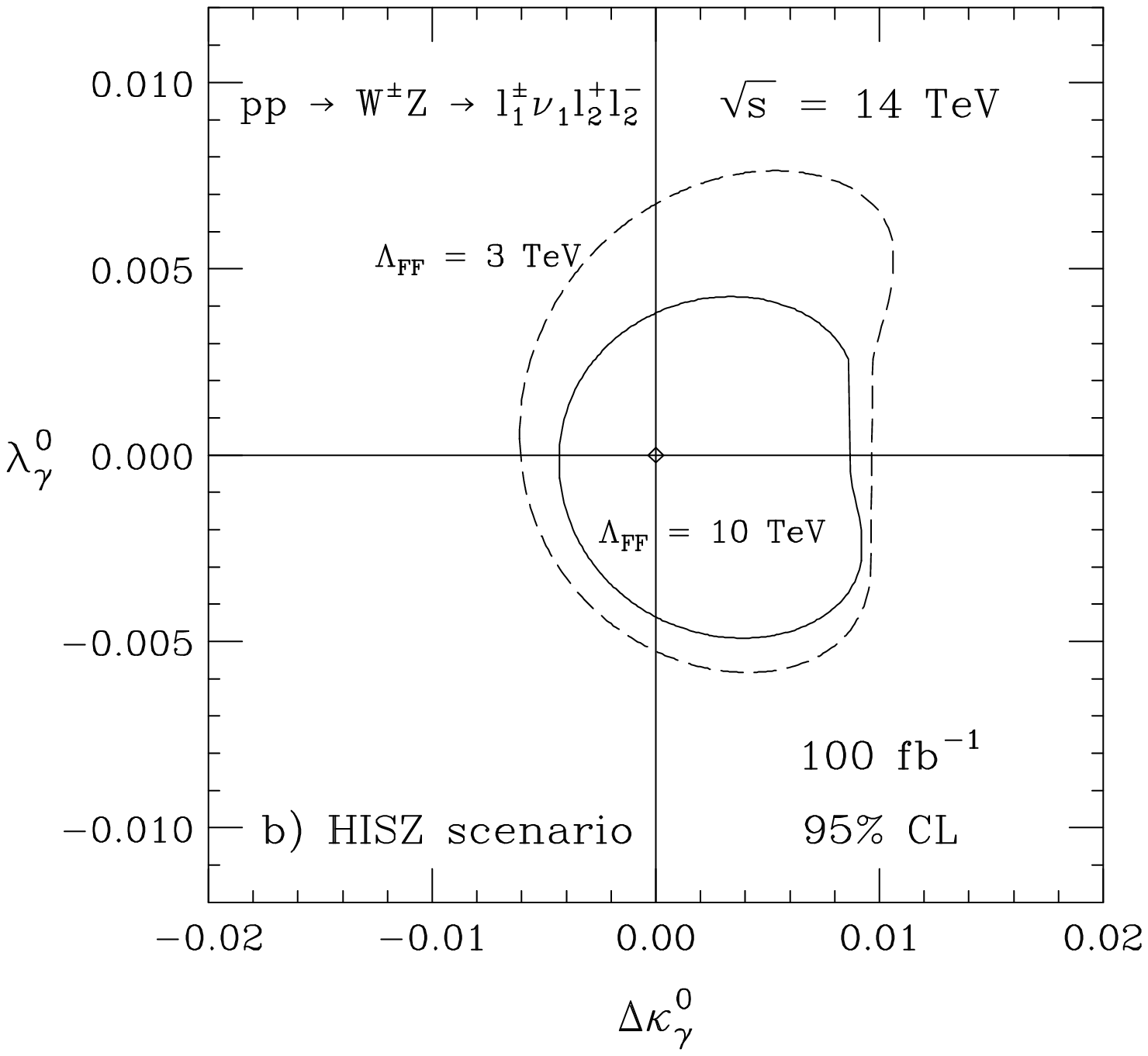,width=2.75in} \\
%\vspace{20pt}
\end{tabular}
\caption{Expected 95\% CL limits on non-standard $WWV$ couplings from
future hadron collider experiments. Part a) shows the projected
sensitivity for $W\gamma$ production at the Tevatron. Part b) displays
the limits one expects from $WZ$ production at the LHC in the HISZ
scenario~\protect{[\ref{hisz1}]}, where $\Delta 
g_1^Z=\Delta\kappa_\gamma/2\cos^2\theta_W$,
$\Delta\kappa_Z=(1-\tan^2\theta_W)\Delta\kappa_\gamma/2$, and
$\lambda_Z=\lambda_\gamma$.}
\label{FIG:TWO}
\end{figure}

In run~2 at the Tevatron, integrated luminosities of at least
1~fb$^{-1}$ per experiment are foreseen. Through further upgrades in the
Tevatron
accelerator complex, an additional factor~10 in integrated luminosity
may be gained (TeV33). With 1~fb$^{-1}$ (10~fb$^{-1}$) one expects
to improve the present limits on anomalous couplings by about a factor~3
(5). As an example, Fig.~\ref{FIG:TWO}a shows the 95\% CL limits expected from
$W(\to e\nu)\gamma$ production for 1~fb$^{-1}$ and 10~fb$^{-1}$~\cite{aihara}.

At the LHC, sensitivities of ${\cal O}(10^{-2})$ can be
reached~\cite{aihara} with an integrated luminosity of
100~fb$^{-1}$. This is illustrated in Fig.~\ref{FIG:TWO}b where we show
the limits which can be reached in $pp\to W^\pm
Z\to\ell_1^\pm\nu_1\ell_2^+\ell_2^-$ ($\ell_1,\,\ell_2=e,\,\mu$). The 
sensitivity bounds which can
be obtained at the LHC depend significantly on the form factor scale. 

\section*{Measuring $WWV$ Couplings at the NLC and a $\mu^+\mu^-$
Collider}
Since the LEP2 center of mass energy is only slightly above the $W$
pair threshold, the SM gauge cancellations are not fully operative, and the
sensitivity to anomalous gauge boson couplings is limited. Much better
limits on $WWV$ couplings will be possible at a linear $e^+e^-$
collider (NLC), or a $\mu^+\mu^-$ collider (FMC), operating in the
several hundred GeV range or above. 

A variety of processes can be used to constrain the vector boson
self-interactions at the NLC or FMC. Because the limits obtained from 
$W$ pair production~\cite{NLCWW} are comparable or better 
than those obtained from other processes, we restrict 
ourselves to this process in the following. 

A study~\cite{tim} based on ideal reconstruction of $W$ daughter pairs
in $e^+e^-\to W^+W^-\to\ell\nu jj$ and ignoring initial state radiation 
(ISR), found the following 95\% CL limits in the HISZ scenario~\cite{hisz} 
($\sqrt{s}=500$~GeV, $\int\!{\cal L}dt=80~{\rm fb}^{-1}$): 
\begin{equation}
|\Delta\kappa_\gamma|<0.0024 \quad (\lambda_\gamma=0), \quad\qquad
|\lambda_\gamma|<0.0018 \quad (\Delta\kappa_\gamma=0).
\end{equation}
To ensure that events are well within the detector volume, and to
suppress the contribution from the $t$-channel $\nu$-exchange diagram, a
$|\cos\Theta_W|<0.8$ cut on the $W$ production angle $\Theta_W$ has been
imposed. 

No detailed simulations have yet been carried out for the FMC.
Nevertheless, a few general conclusions can be
drawn. The main differences between $W$ pair production in $e^+e^-$ and
$\mu^+\mu^-$ collisions are
\begin{itemize}
\item the background in the detector caused by the decay of the
muons~\cite{geer}, and 
\item the reduced level of initial state radiation due to the larger
mass of the muons. 
\end{itemize}
In the following we concentrate on the impact of initial state radiation
on the measurement of the $WWV$ couplings at a lepton collider. 

For the NLC, these effects were studied in Refs.~\cite{keith}
and~\cite{hansen}. Initial state radiation is strongly peaked in the
beam direction and at zero photon energy. Many photons originating from
ISR therefore are not detected. The emission of an undetected 
photon along the beam
direction affects the kinematics of the visible decay products, and a
kinematic fit assuming full energy momentum conservation results in
incorrect production and decay angles. In addition, photon emission
leads to a reduced effective center of mass energy, and thus to
distorted angular distributions. 

\begin{figure}[t] % fig 3
\centerline{\epsfig{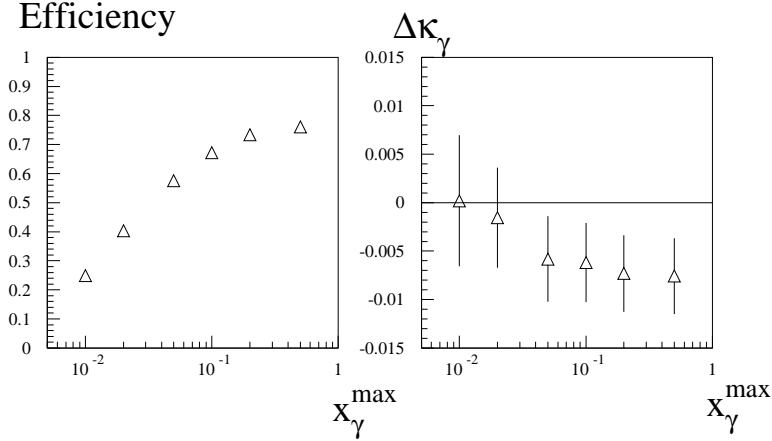}}
%\vspace{20pt}
\caption{Efficiency and bias for $\Delta\kappa_\gamma$ in the HISZ
scenario as a function of the maximum allowed fraction of center of mass
energy for the {\sl fitted} energy of a photon undetected along the beam
pipe in $e^+e^-\to W^+W^-\to\ell\nu jj$ ($\sqrt{s}=500$~GeV;
$\int\!{\cal L}dt=10~{\rm fb}^{-1}$). The error bars indicate the
statistical error in $\Delta\kappa_\gamma$.}
\label{FIG:THREE}
\end{figure}

In $WW\to jjjj$ events, the photon momentum vector can in principle be fully 
reconstructed. In the semileptonic case, $WW\to\ell\nu jj$, the presence
of the neutrino makes this more difficult. Assuming that the photon is
emitted in beam direction, a kinematic fit to the event can be
performed. ISR effects can be reduced by requiring that the fitted photon
energy is less than a fraction $x_\gamma^{max}$ of the nominal center of mass
energy. Figure~\ref{FIG:THREE}a shows the efficiencies obtained for
$\sqrt{s}=500$~GeV and $\int\!{\cal L}dt=10~{\rm fb}^{-1}$ as a function
of $x_\gamma^{max}$~\cite{keith}. The presence of the undetected
neutrino leads to a
degradation in the photon energy resolution, resulting in a large loss
in efficiency for small values of $x_\gamma^{max}$. 
Figure~\ref{FIG:THREE}b displays the corresponding biases in fitted
$\Delta\kappa_\gamma$ values for the HISZ scenario as a function of
$x_\gamma^{max}$. The bias is small compared to the statistical error
only for $x_\gamma^{max}<0.02$. Due to the loss in efficiency at small 
$x_\gamma^{max}$, the error on $\Delta\kappa_\gamma$ increases
significantly with decreasing $x_\gamma^{max}$. 

For $WW\to\ell\nu\ell'\nu'$ events the momenta of the two neutrinos are
unknown. If ISR is ignored, the neutrino momenta can be reconstructed up
to a two-fold ambiguity. Initial state radiation influences the
existence of solutions, and introduces a large bias in the extracted
values of anomalous couplings~\cite{hansen}. 

At the FMC, one expects the effects caused by ISR to be
substantially smaller than at the NLC. It should therefore
be possible to measure anomalous $WWV$ couplings at a $\mu^+\mu^-$ 
collider with similar or better precision than at a $e^+e^-$ machine 
operating at the same center of mass energy and luminosity, unless
backgrounds from muon decay in the detector play an important role. Clearly,
more detailed work is needed before definite conclusions can be drawn.

\section*{Conclusions}
Within the past few years, our experimental knowledge of the gauge boson
self-interactions has grown very rapidly. The $WWV$ coupling parameters
have been measured with an accuracy of 20~--~45\% at the
Tevatron. Within the next few years the limits on anomalous couplings
are expected to improve by a factor 4~to~5 by experiments at LEP2 and
the Tevatron. At the LHC, one hopes to probe non-standard $WWV$
couplings with a precision of ${\cal O}(10^{-2})$. At the NLC, anomalous
couplings can be tested at the ${\cal O}(10^{-3})$ level. A similar or
better precision is expected for a $\mu^+\mu^-$ collider operating at
the same energy and luminosity.

\end{document}